\documentclass{aastex631}

\newcommand\dv{\Delta v}
\newcommand\db{\Delta b}
\newcommand\dtheta{\Delta \theta}

\def\las{\mathrel{\hbox{\rlap{\hbox{\lower3pt\hbox{$\sim$}}}\hbox{\raise2pt\hbox{$<$}}}}}
\def\gas{\mathrel{\hbox{\rlap{\hbox{\lower3pt\hbox{$\sim$}}}\hbox{\raise2pt\hbox{$>$}}}}}

\begin{document}

\author[0000-0002-1914-5352]{Paul Wiegert}
\affiliation{Department of Physics and Astronomy\\The University of Western Ontario\\London Ontario Canada}
\affiliation{Institute for Earth and Space Exploration \\The University of Western Ontario\\London Ontario Canada}

\title{On the sensitivity of Apophis' 2029 Earth approach to small asteroid impacts}

\begin{abstract}

Apophis' current trajectory takes it safely past our planet at a
distance of several Earth radii on 2029 April 13. Here the possibility
is considered that Apophis could collide with a small asteroid, like
the ones that frequently and unpredictably strike Earth, and the
resulting perturbation of its trajectory. The probability of an
impact that could significantly displace Apophis relative to its
keyholes is found to be less than 1 in $10^6$, requiring a $\dv \gas
0.3$mm/s, while for an impact that could significantly displace
Apophis compared to its miss distance in 2029 it is less than 1 in
$10^9$, requiring a $\dv \gas 5$cm/s. These probabilities are below
the usual thresholds considered by asteroid impact warning systems.

Apophis is in the daytime sky and unobservable from mid-2021 to
2027. It will be challenging to determine from single night
observations in 2027 if Apophis has moved on the target plane enough
to enter a dangerous keyhole, as the deviation from the nominal
ephemeris might be only a few tenths of an arcsecond. An impending
Earth impact would, however, be signalled clearly in most cases by
deviations of tens of arcseconds of Apophis from its nominal ephemeris
in 2027. Thus most of the impact risk could be retired by a single
observation of Apophis in 2027, though a minority of cases present
some ambiguity and are discussed in more detail. Charts of the on-sky
position of Apophis under different scenarios are presented for
quick assessment by observers.

\end{abstract}



\section{Introduction} \label{sec:intro}

Asteroid 99942 Apophis has been among the most heavily-studied
near-Earth asteroids (NEAs) because of its much-anticipated close
approach to Earth in 2029. Careful observation of this body since its
discovery in 2004, and painstaking computation of its future path
under the laws of physics show it missing our planet by a safe margin
on 2029 April 13\footnote{\url{https://www.jpl.nasa.gov/news/nasa-analysis-earth-is-safe-from-asteroid-apophis-for-100-plus-years} \label{JPLpressrelease},
retrieved 2024 April 15}. However even the best current projections do not account for the
possibility of Apophis being struck by a small asteroid, such as
regularly hit Earth and which are often widely visible in the form of
meteors and fireballs. The impact of a small asteroid on Apophis
---though very unlikely--- would create a small impulsive change in
the velocity of that asteroid. Such an asteroid strike on Apophis would be
  the natural analog of the kinetic impact process by which a
$2.70\pm0.10 \times 10^{-3}$ m/s velocity change was imparted to
asteroid Dimorphos by the Double Asteroid Redirection Test (DART)
spacecraft in 2021 \citep{cheagrbar23}.

In this work we will show that the probability of an impact sufficient
to deflect Apophis in a dangerous manner is exceedingly low. However,
the fact that Apophis' calculated path brings it near Earth at a
distance of only $38012\pm 4$ km (center-to-center) on 2029
  April
  13\footnote{\url{https://ssd.jpl.nasa.gov/tools/sbdb_lookup.html\#/?sstr=99942}\label{JPLdetails},
retrieved 2024 April 15} drives us to consider even such highly
improbable cases, to ensure our collective understanding of this
asteroid's future motion is as complete as possible.

Asteroid collisions with other asteroids or comets are rare, but
have been observed several times.  Asteroid 596 Scheila, comet
354P/LINEAR and asteroid 493 Griseldis are all thought to have been
struck by small asteroids
\citep{ishhanhas11,jewweamut11,snotubvin10,jewweaaga10,thoshetru15}. Asteroid
families are also believed to be the result of collisions
\citep{micbentan01}.

\cite{wiehya24} showed that there are no expected collisions of
Apophis with the catalogue of 1.2 million known asteroids, though
there is some very low chance of collision with material accompanying
a few of them. However, it is not known at this time whether the
asteroids in question even have accompanying material. The possibility
of an impact with material accompanying a known asteroid is implicitly
included in our analysis here, but we are primarily concerned with the
background population of small unseen meter-class asteroids and
meteoroids that could hit Apophis at any time.

Though the calculated path of Apophis during its 2029 return is
confidently expected to keep it from impacting our planet, the
question of whether or not it will pass through a keyhole is a more
complicated problem.  A keyhole refers to a region of space near Earth
which, if Apophis should pass through it during its 2029 close
approach, indicates that it is on a collision course with Earth at a
future date. Much work and attention was given to the keyhole problem,
and in 2021, the problem was retired, at least for keyholes leading to
impacts up to and including 2068, by a careful analysis from JPL$^{\ref{JPLpressrelease}}$. This
analysis is based on a careful reconstruction of Apophis' path from
telescopic and radar observations. However it does not include the
possibility of a random impulsive change in Apophis' velocity such as
might occur from a collision with a small asteroid.

An additional element of the story is that Apophis has been largely
unmonitored by telescopes since May 2021, and will remain so through
2027. This simply arises because of the relative geometry of Apophis,
Earth and the Sun, which puts the asteroid in the daytime sky for the
time span in question. Even professional telescopes cannot observe
Apophis in broad daylight, in the same way that one cannot see the
stars with the unaided eye during the day. It does not result from a
defect in the eye or in the telescope, but simply from the
overwhelming brightness of the daytime sky. Most space telescopes also
avoid looking towards the Sun; the Hubble Space Telescope has a solar
avoidance zone of $55\deg$ \citep{HST2024} and the James Webb Space
Telescope, $85\deg$ \citep{nelatcatk04}. The Solar and Heliospheric
Observatory (SOHO) can and does observe near the Sun, but its Large
Angle Spectrometric Coronagraph (LASCO) has a limiting magnitude of
$V=9.5$ for solar system objects \citep{lamfaulle13} while Apophis is
fainter than $V=19$ until 2027.  As a result, Apophis will remain
essentially unmonitored for over six years. If it did suffer a
significant impact during this time we might be unaware of the fact,
and only able to infer that a perturbation occurred from observations
taken once it returns to observability in 2027.

The purpose of this paper is to model the effect of a small asteroid
impact on Apophis occurring after observations of it were forced to
cease, in order to determine:
 \begin{enumerate}
  \item What magnitude of velocity change would be needed to significantly affect the details of Apophis' 2029 close approach? That is, what impulse could move Apophis on the target plane by an amount that could direct it into a keyhole to a collision at a future date, or to collision with the Earth or Moon in 2029? 

  \item What size asteroid would need to strike Apophis to create such an impulse, and what is the likelihood of such a collision occurring?

\item If Apophis suffered such an impact, how would the on-sky position of Apophis differ from its nominal ephemeris when it re-emerges to visibility in 2027? If Apophis' trajectory is found to differ from that expected, how can we quickly identify dangerous trajectories?
  
\end{enumerate}

The probability of a small asteroid colliding with any asteroid over
the course of a few years or decades is tiny and indeed could be
considered negligibly small in almost any other case. It is only this
asteroid's unusual trajectory and the attendant risk it poses to our
planet that drives us to examine this scenario in detail here.

 \section{Methods} \label{sec:methods}

\paragraph{Time frame of the study}
We begin our study at the earliest time at which Apophis could have
had an undetected impact, set by the last observations of Apophis
recorded before it moved too close to the Sun on the sky. These last
observations represent the final times when the presence of dust from
a hypervelocity impact onto Apophis could have been detected
telescopically. At this writing, the Minor Planet Center shows the
last few observations of Apophis occur in May 2021 through April
2022\footnote{\url{https://minorplanetcenter.net/db_search/show_object?object_id=Apophis}, \label{MPCobservations}
retrieved 2023 April 13}.  The last observation from the major NEA
surveys was 2021 May 12 for both PanSTARRS 1 Haleakala and Steward
Observatory, Kitt Peak-Spacewatch, observations taken at a solar
elongation angle of 63 degrees \citep{mps1406942}.  Observatory B18 -
Terskol in Ukraine does report a additional position on 2021 May 20 at
an elongation of 53 degrees \citep{mps1480581} and there is an
occultation observation listed in April 2022 \citep{mpec2024-c06}.

The major NEA surveys consistently provide the most reliable observations
of near-Earth objects.  And given that the DART impact into Dimorphos
caused a noticeable dust cloud and brightening within minutes to hours
after the event \citep{karthoyan23}, we can expect that a sizable
impact on Apophis would be almost immediately detected by the major
surveys if observing conditions were good.

The observations taken of Apophis after those of the major surveys
represent important contributions to the monitoring of Apophis, but it
is unclear whether they would have detected dust near Apophis under
difficult near-Sun viewing. Occultation observations, often collected
by amateur astronomers under challenging conditions, represent
valuable high-quality measurements of Apophis' on-sky position but are
not optimized for dust detection. As a result, we will adopt the date
after the final observation by the major NEA surveys, 2021 May 13, as
the first date at which an unseen impact might have occurred. We will
see that earlier impulses do as a rule produce larger deviations from
Apophis' expected trajectory, but that our results are not very
sensitive to the precise timing of the impulse.

Based on the final recorded observations of Apophis, we will also
adopt a solar elongation of 60 degrees as the practical though not
absolute limit at which Apophis observations can be made. We will use
this value to determine when Apophis can be re-observed again when it
re-emerges from the daytime sky. We note that Apophis returned to
solar elongations greater than 60 degrees during October 2021 to April 2022
at $m_V \approx 21$ but its solar elongation did not exceed 80
degrees. During this time, only the occultation observation mentioned
earlier is recorded by the Minor Planet
Center$^{\ref{MPCobservations}}$.

\paragraph{Model}

We compute the path of Apophis within a model solar system, described
below. To examine the effects of an asteroid impact on Apophis, we
create a number of hypothetical versions of Apophis (``clones''). Each
clone has the same initial conditions as Apophis at the start of the
simulation, but each clone has a single velocity impulse applied to
it. The impulse has a fixed magnitude $\dv$ which we assume is
uniformly distributed on the sphere, though in reality some
directional asymmetry is expected \citep{lefwie08, robpokgra21}. In
practice, the impulse direction is chosen by selecting three random
numbers ($R_x$, $R_y$, $R_z$) uniformly within the interval (-1,
+1). If $R_x^2 + R_y^2 + R_z^2 > 1$, the numbers are discarded and new
ones drawn. If $R_x^2 + R_y^2 + R_z^2 <= 1$, then ($R_x$, $R_y$,
$R_z$) is normalized and defines impulse direction of $\dv$ in
heliocentric Cartesian coordinates. This process ensures the impulses
are indeed uniformly distributed on the sphere. The impulse occurs at
a time chosen at a uniformly random point within the time frame
examined. This time frame extends from 2021 May 13 (JD 2459347.5)
through to Apophis' next close approach on 2029 April 13 (JD
2462239.5).

Apophis and its clones are modelled within a solar system which
includes the eight planets and the Moon. Planetary initial conditions
are from the JPL DE405 ephemeris \citep{sta98}. All particles are
integrated with the RADAU \citep{eve79} algorithm with a tolerance of
$10^{-12}$.  Post-Newtonian effects are included, as is the Yarkovsky
effect with parameters as determined by JPL and which are assumed to be
unchanged by the small asteroid impact. Apophis and its clones are
treated as massless test particles.

\paragraph{Model tests} For testing purposes, we propagated our model
forward eight years from our initial conditions to the close approach
to Earth on 2029 April 13. Our model reports a close approach distance
of 38105~km, 93~km ($< 10^{-8}$~au) from JPL CNEOS's current value of
$38012\pm4$~km$^{\ref{JPLdetails}}$, and occurring within 10 seconds
of the time reported by JPL.  Our model therefore matches the JPL
values closely.  JPL does not report the target $b$-plane position
associated with their current orbital solution for Apophis, but our
$b$-plane error is certainly similar to that in the close approach
distance, about 100~km. We will see that this is sufficient to resolve
changes of interest (which will be at their smallest about 200 km on
the b-plane, see Section~\ref{sec:sensitivity}).  Our uncertainties
are somewhat larger than JPL's but sufficiently accurate for the
sensitivity study presented in this paper.

\section{Results and discussion}

\subsection{Sensitivity to impulses} \label{sec:sensitivity}

Could the impulsive change to Apophis' velocity caused by a small
asteroid impact push Apophis into a keyhole? The keyhole analysis of
\cite{farchecho13} reveals that the pattern of keyholes on the target
plane is very complex. Our goal here is not to determine the
circumstances that would put Apophis into a specific keyhole. Rather
we simply ask what size impulse would be needed to move Apophis
substantially with respect to the most important keyholes, and thus
require a possible re-evaluation of its future impact probability.

\cite{farchecho13} reveals that Apophis would have to move
approximately 200~km vertically on the target plane to reach the
nearest important keyholes, and about 1500~km vertically on the target
plane to reach the complex of secondary keyholes surrounding the 2036
primary resonant return. We will adopt these numbers as representative
of the target plane displacements of interest.

\begin{figure}[ht!]
\plottwo{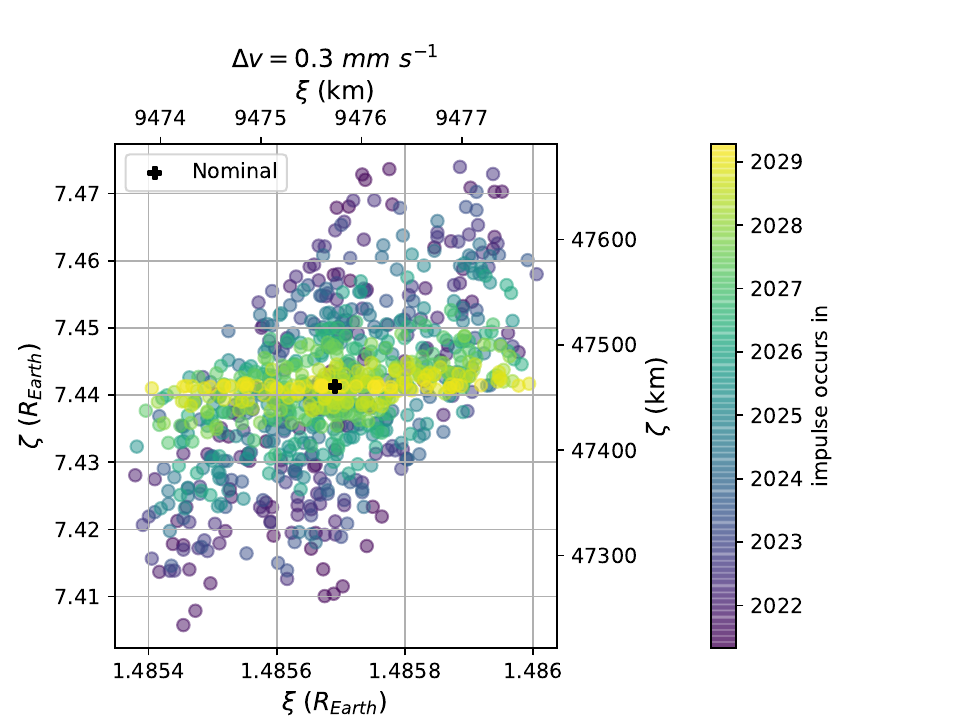}{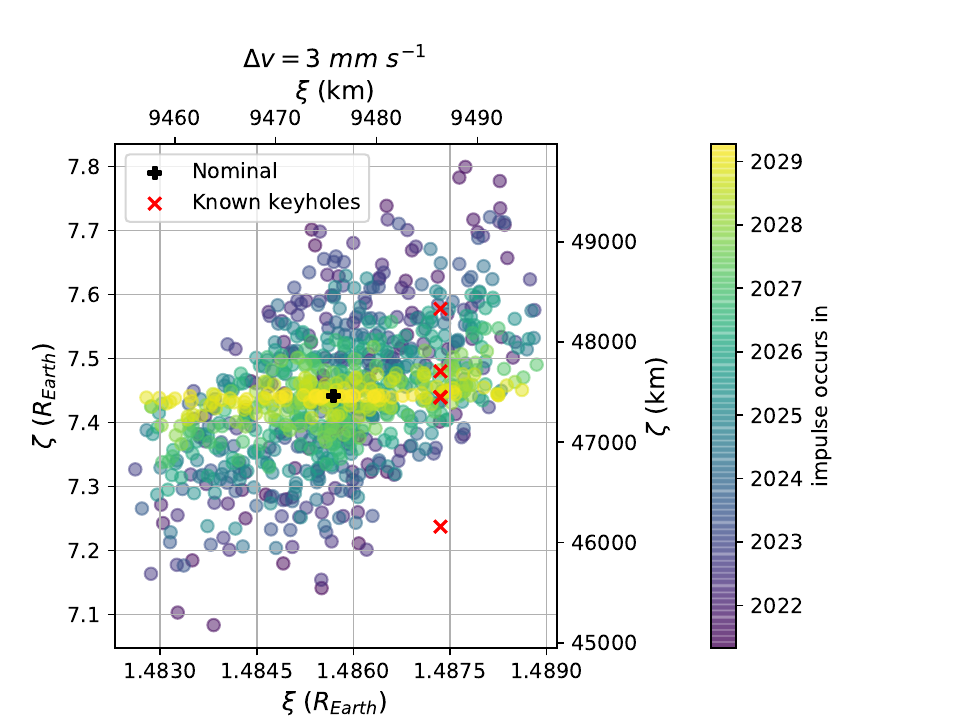}
\caption{The Opik target plane under $\dv = 3 \times 10^{-4}$~m/s
  (left) and $3 \times 10^{-3}$~m/s (right) impulses. Impulses of this
  order are able to produce displacements of Apophis on the target
  plane of $\sim200$~km and $\sim1500$~km respectively in the vertical
  direction, necessary to move the asteroid into keyholes. However,
  only very specific (and hence very unlikely) impulses will result in
  keyhole entry. Some important keyholes identified by \cite{farchecho13} are indicated.}
\label{fig:bplane-keyholes}
\end{figure}

\begin{figure}[ht!]
\plotone{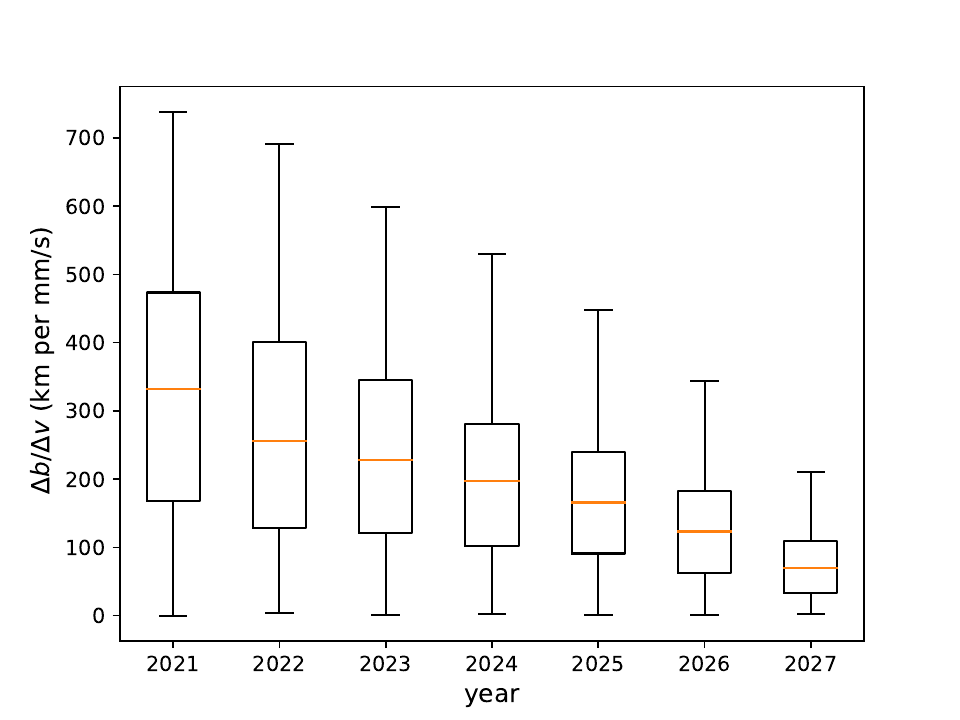}
\caption{A box plot of $\db/\dv$ over time. The boxes show the inner quartiles with a line at the median, while the whiskers extend across the full range of values. Impacts that occur earlier in time are more effective at moving Apophis on the target plane.}
\label{fig:dbdv}
\end{figure}

From our simulations we find that a $\dv$ of $3 \times 10^{-4}$~m/s is
sufficient to move Apophis 200~km on the b-plane to the nearest
keyholes, and $3 \times 10^{-3}$~m/s to move it 1500~km on the b-plane
to the more distant 2036 keyhole complex (see
Fig~\ref{fig:bplane-keyholes}).  Asteroid impacts are found to
move Apophis efficiently vertically on the target plane; however, any
particular impulse is as likely to move Apophis away from a keyhole as
towards it, depending on the direction in which it was applied. A
  plot of the median $\db/\dv$ over time in our simulations is
  presented in Figure~\ref{fig:dbdv}. We conclude that a
$\dv \sim 3 \times 10^{-4}$ m/s is the minimum impulse of concern,
that is, it is the minimum necessary to deflect Apophis to one of the
 nearby keyholes. Impacts occurring later in
  time are less effective in moving Apophis on the target plane. Even
if such an impulse should occur, the results would most likely be
harmless. Nonetheless, monitoring Apophis telescopically for signs of
such an event at the earliest possible date is recommended (see
Section~\ref{sec:onsky}).

\begin{figure}[ht!]
\plottwo{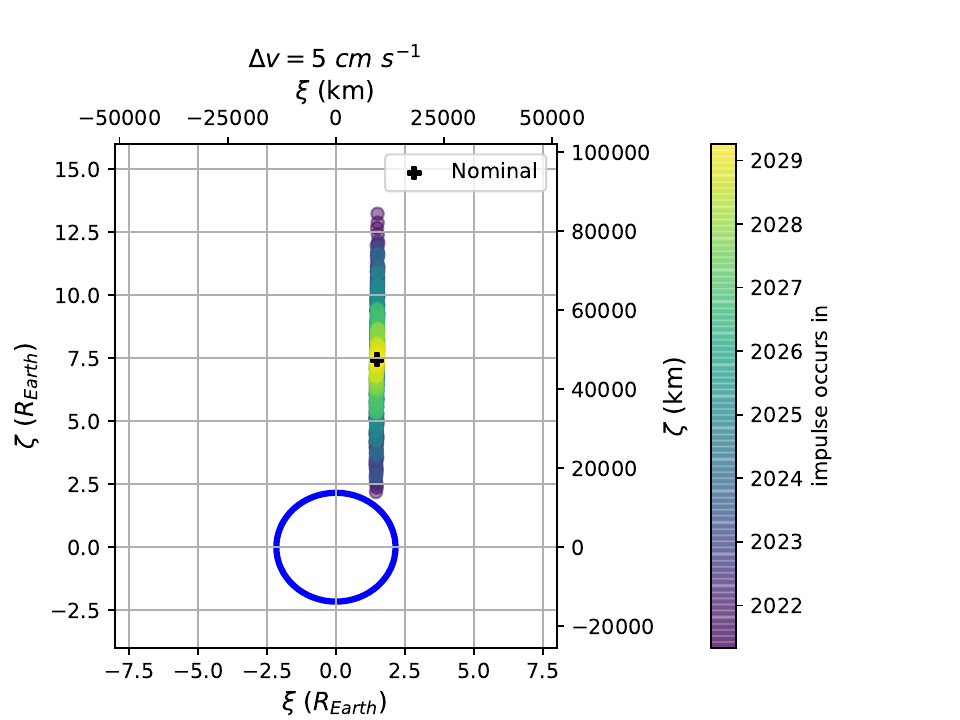}{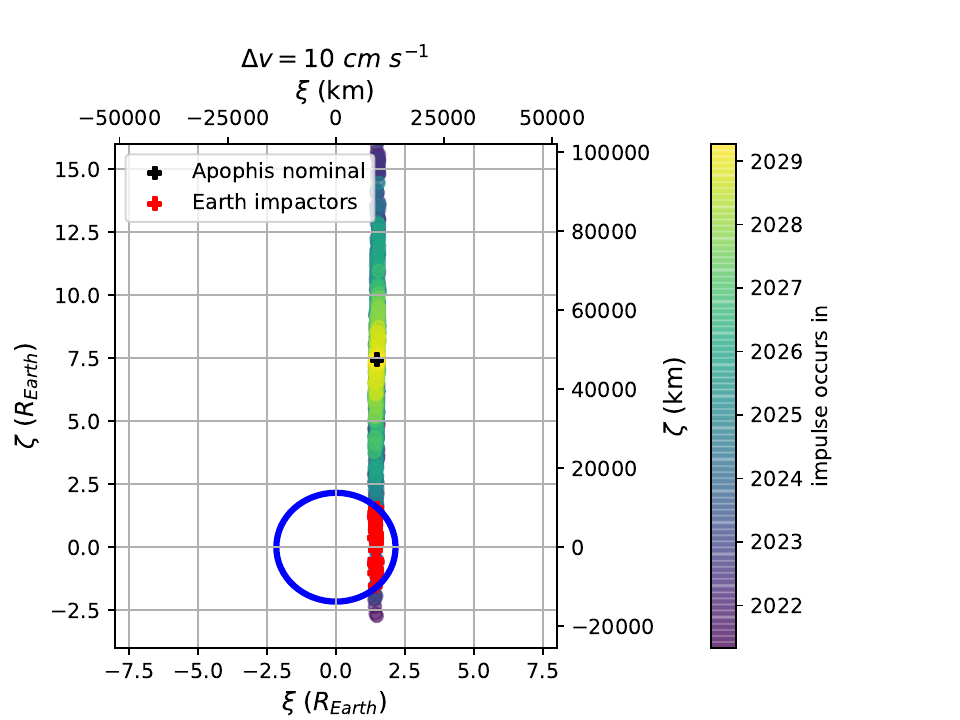}
\caption{The Opik target plane under $5 \times 10^{-2}$~m/s (left) and
  $10^{-1}$~m/s (right) impulses. The impact cross-section of the
  Earth in this case is 2.16~Earth radii \citep{farchecho13}, and is
  indicated by the blue circle. Any point inside the circle represents
  a collision with our planet (red crosses). An impulse of $5 \times
  10^{-2}$~m/s could just reach Earth if applied early and directed in
  just the right direction. Impulses of $\dv \gas 10^{-1}$~m/s may
  reach Earth over a wider range of impulse directions and timings. }
\label{fig:bplane-impact}
\end{figure}

\begin{figure}[ht!]
\plotone{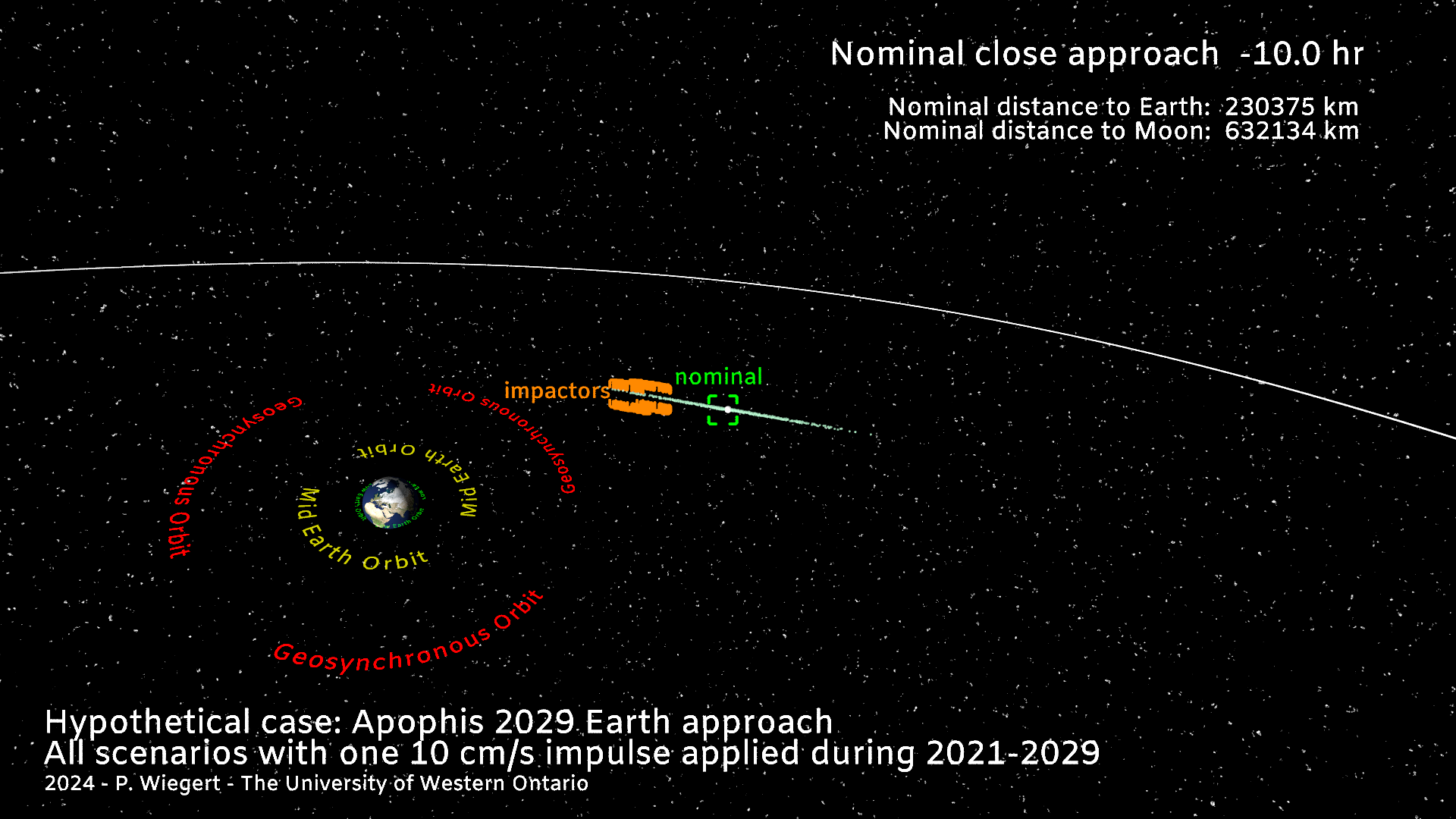}
\caption{Illustration of the cases where one impulse $\dv = 10^{-1}$~m/s
  from a small asteroid impact has been applied to Apophis. The
  nominal (unperturbed) case is highlighted by a green reticle. Cases
  leading to Earth impact are indicated by an orange
  reticle.\label{fig:animation} This figure is a single frame of an animation
  showing the approach of the asteroid and clones to Earth. The animation
  is available in the online journal, or at \url{https://youtu.be/9UYM6MUqvYM}.
  The animation is approximately 20 seconds long, and shows the relative
    positions of the Earth, Moon, Apophis and the clones during several hours around the 2029 close approach.
}
\end{figure}

Our simulations show that small asteroid impacts move Apophis
efficiently in the vertical ($\zeta$) direction on the target
plane. Since the encounter geometry of Apophis with Earth is such that
a sufficient vertical displacement on the target plane would bring
about an impact with our planet, we examine this possibility here as
well. An impulse $\dv$ of $5 \times 10^{-2}$~m/s is found to be the
minimum value that could result in impact with Earth in 2029, though
the impact would have to be applied at a very specific time and
direction for this to occur (see Fig~\ref{fig:bplane-impact}). At $\dv
\gas 10^{-1}$~m/s, a wider range of impulse directions and timings can
produce an Earth impact. An animation of the scenario with $\dv =
10$~cm/s is shown in Figure~\ref{fig:animation}.
For impulses in the 0.1-1~m/s range (the largest impulses
considered here), the resulting probability of Earth impact hovers
near 5\%, as only certain impulse directions can direct Apophis to our
planet, and larger impulses eventually start to miss Earth by
passing on the opposite side (see Fig~\ref{fig:bplane-impact}).

Small asteroid impacts cannot easily push Apophis into a collision
with the Moon in 2029. The nominal close approach distance of Apophis
to the Moon is about 96,000km (55 $R_{Moon}$). None of the cases examined
here approach the Moon more closely than 40 $R_{Moon}$, all remain
outside the Moon's sphere of influence (37 $R_{Moon}$).

Having determined the magnitudes $\dv$ of the impulses of interest, we
now turn to an analysis of the probability of an asteroid of
the necessary size striking Apophis.

\subsection{Probability of small asteroid impacts on Apophis} \label{sec:probability}

What size asteroid would need to hit Apophis in order to generate
the impulses described earlier? To an order of magnitude, the change
in velocity $\dv$ in Apophis due to an asteroid impact scales by the
impactor's relative momentum
\begin{eqnarray}
   \beta m v  &\sim& M\dv \nonumber\\ 
  \dv&\sim& \frac{\beta m v}{M} \label{eq:dv}
\end{eqnarray}
where $M$ is Apophis' mass, $m \ll M$ is the impactor mass, $v$ is the
speed of the impactor relative to the asteroid, and $\beta$ is an
enhancement factor resulting from momentum carried away by ejecta.
For context we consider the measured $\dv$ of the DART impact into
Dimorphos. Dimorphos has a 151~m diameter and a mass $M=4.3 \times
10^9$ kg assuming a similar density to Didymos, while the DART
impactor had a mass of 580~kg, hit at a relative speed of 6.1~km/s and
had a measured $\beta$ between 2 and 5 \citep{dalernbar23}. Adopting
$\beta=3$, Eqn~\ref{eq:dv} yields $\dv = 2.4 \times 10^{-3}$m/s, while
the measured velocity change of the DART impact was of $2.70\pm0.10
\times 10^{-3}$ m/s \citep{cheagrbar23}. \cite{cheagrbar23} find a
mean value of $\beta=3.6$ which yields $\dv = 2.88 \times 10^{-3}$m/s.
These close matches reveal that Eq~\ref{eq:dv} is sufficient for our
purposes.

The diameter $d$ needed for an impacting asteroid to create a
particular $\dv$ in Apophis can be found from Eq~\ref{eq:dv}, assuming
roughly spherical shape and density $\rho$, as
\begin{equation}
d \sim  9.2  \left(  \frac{\dv}{\rm 1~m/s} \right)^{\frac{1}{3}} \left( \frac{\beta}{3}\right)^{-\frac{1}{3}} \left( \frac{\rho}{3500~\rm{kg~m}^3} \right) ^{-1/3} \left( \frac{v}{17 \rm{~km/s}} \right)^{-\frac{1}{3}} \left( \frac{D}{{\rm 340~m}} \right) \rm{m} \label{eq:diameter}
\end{equation}
where $D=340$~m is Apophis' diameter \citep{brobenmcm18}, and its
density is taken to be 3500~kg~m$^3$ \citep{binrivtho09}. The average
impact velocity of bolides arriving at Earth $v\approx 20$~km/s
\citep{gruzoofec85, brosparev02, grengogla12,droottkos20} which
includes a 11 km/s component due to Earth's gravitational attraction
which will not apply at Apophis, so we adopt $v \approx \sqrt{20^2
  -11^2} \approx 17$~km/s for the typical relative velocity.

Equation~\ref{eq:diameter} reveals that an impact able to
move Apophis significantly relative to the 2029 keyholes ($\dv \gas
3 \times 10^{-4}$m/s) would necessitate an impactor with $d \gas 0.6$~m. In
order to create the possibility of an impact with Earth itself in
2029, a $\dv \gas 5 \times 10^{-2}$~m/s is needed, translating to 
 $d \gas 3.4$~m.

What are the probabilities of asteroids of sufficient size striking
Apophis?  Adopting the measured flux of meter-sized objects at Earth
from \cite{brosparev02} we see that approximately 140 0.6-m diameter
or larger objects, and only a single 3.4 m or larger object, strike
the Earth each year on average. However, Apophis's physical
cross-section is only (0.17 km/6378 km)$^2\ \approx 7 \times 10^{-10}$
that of Earth, so the rate of impacts is reduced by this factor. This
puts the odds of Apophis being struck by an asteroid large enough to
deflect it significantly relative to the 2029 keyholes is less than 1
in one million. Given that not all impulses of a given size move
Apophis towards a keyhole, the odds of a small asteroid impact
deflecting Apophis to a dangerous post-2029 encounter over the 8 year
time frame examined here is minuscule. The odds of an unseen small
asteroid deflecting Apophis enough to direct it into a collision with
Earth in 2029 ($d \gas 3.4$m, $\dv > 5 \times 10^{-2}$~m/s) is
approximately $10^{-8}$.  Given that only 5\% of such impulses are in
the correct direction to generate an Earth impact (see
\ref{sec:sensitivity}), the overall probability of a small impact
directing Apophis into collision with the Earth is less than one in 2
billion.

Our results depend on the densities $\rho$ and velocities $v$ of the
small impactors at a given size. If either of these is higher than we
assumed here, then the resulting $\dv$ at a given diameter is also
increased, and smaller impactors could create the necessary
impulses. This could increase the probability of the deflections
considered here occurring, since smaller impactors are more
abundant. However, Eqn~\ref{eq:diameter} depends on $v$ and $\rho$
only to the negative one-third power, and so the probabilities
discussed here are not very sensitive to the details of the impactor
population.

\subsection{On-sky position of Apophis} \label{sec:onsky}

We showed that the probability of a small asteroid deflecting
Apophis substantially is exceedingly small. For any other asteroid,
the eventuality might be dismissed as negligible. However given the
disproportionate risks associated with Apophis, we turn to an
examination of how and when we could determine whether such a deflection
has occurred.

Apophis is as of this writing in the daytime sky, and cannot be
monitored telescopically. If Apophis suffers an impact before it
returns to visibility in 2027, the most easily detectable
after-effects (i.e. optical brightening due the impact-produced dust
cloud) could have dissipated. If that is the case, the effect of a
small asteroid impact would perhaps only be revealed to telescopes on
Earth by a deviation of Apophis from its expected position in the
sky. Changes in Apophis' expected magnitude or even its distance could also be
useful; we'll discuss those later in Section~\ref{sec:magpar}.

The detailed determination of an asteroid's trajectory requires large
numbers of careful telescopic observations taken over a period of
weeks, months or even years. Though this process will certainly be
initiated by astronomers around the globe as soon as Apophis
reappears, it is very hard to predict how long a full computation of
Apophis' path based solely on post-2027 observations will
take. Instead, we ask a simpler question here: if Apophis re-emerges
into the night time sky on a trajectory perturbed enough to move it
into an important keyhole or even to a 2029 Earth impact, how would
its on-sky position differ from its current nominal ephemeris?  What
observations could quickly reveal whether it has been significantly
perturbed?

\begin{figure}[ht!]
\plotone{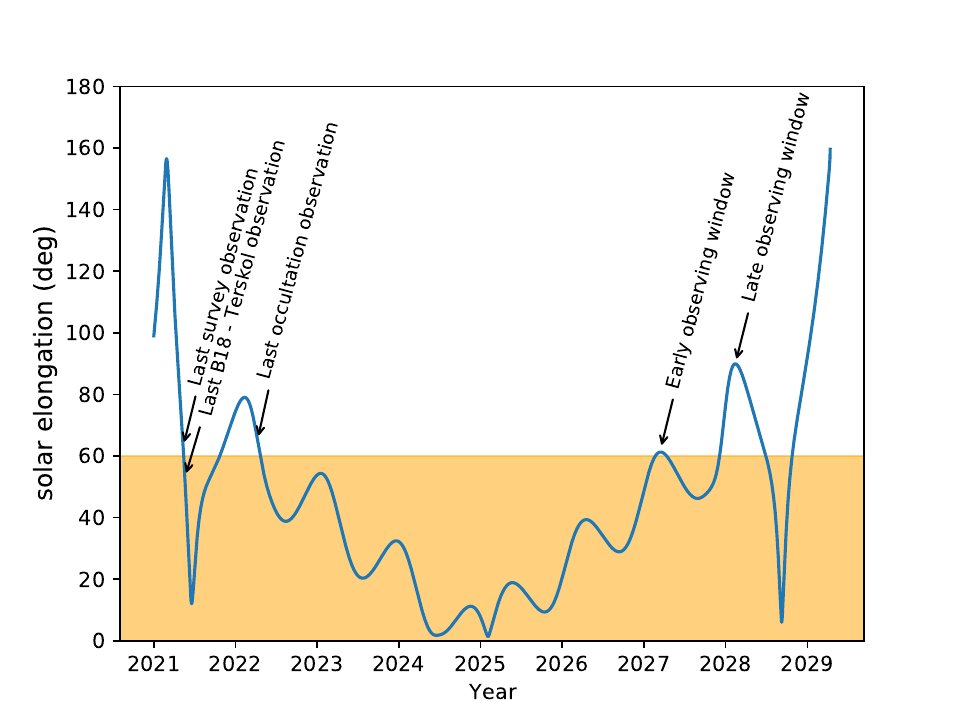}
\caption{The solar elongation of Apophis as seen from Earth during 2021 to 2029. The start  of each year is labelled on the $x$-axis.}
\label{fig:elongation}
\end{figure}

For Earth-based observers, Apophis will emerge ---barely--- from the
solar glare in late February 2027, after having been largely unobservable
for over six years (Figure~\ref{fig:elongation}).  The elongation of
Apophis from the Sun reaches 60 degrees as seen from Earth's geocenter
on 2027 February 22 (JD 2461458.5)\footnote{The observational circumstances
of Apophis reported here are from the JPL Solar System Dynamics Group, NASA/JPL Horizons On-Line Ephemeris System
(Giorgini et al 1996) \url{https://ssd.jpl.nasa.gov/horizons}, \label{Horizons} retrieved 2024 March 17\nocite{gioyeocha96}}. At this point, the asteroid has an expected apparent
magnitude of 21.0. But Apophis does not reach elongations above 62
degrees, hovering near 60 degrees until April 2027 before moving further into
the daytime sky again. Apophis finally re-emerges to elongations
greater than 60 degrees on 2027 December 6 (JD 2461745.5) at magnitude
19.7, remaining visible until 2028 June 30 (2461952.5). These time
ranges will be used here to define ``early'' and ``late'' observing
windows for Apophis.

What deviation from the nominal ephemeris would be detectable? For a
single night's observation from a single station, we will consider
that a displacement of more than 1 arcsec on the sky from the nominal
RA or Dec will be detectable by ground-based telescopes, as this level
is routinely achieved at professional observatories. On-sky positions
can be measured to better than 1~arcsecond under good conditions, but
given that we are considering observations of Apophis taken near the
Sun, with higher than usual scattered light and potentially fewer
visible stars for astrometric solutions, it is a reasonable if
conservative assumption.

In Section~\ref{sec:sensitivity}, we saw that \cite{farchecho13}
determined Apophis will intersect the target $b$-plane 200-1500~km
from the most important keyholes. As a result, we will take those
particles found to move on the $b$-plane by $\db >$~200~km to
represent those which move appreciably with respect to the
keyholes. These cases do not necessarily represent particles which
have been perturbed into a dangerous keyhole. In fact the odds are
very low that keyholes will be entered even under a large
perturbation, but such a determination will require a full orbital
analysis after Apophis reappears in 2027. Our goal here is simply to
provide a framework in which cases of concern can be quickly
identified for further study.

\subsubsection{Keyholes}

\begin{figure}[ht!]
\plottwo{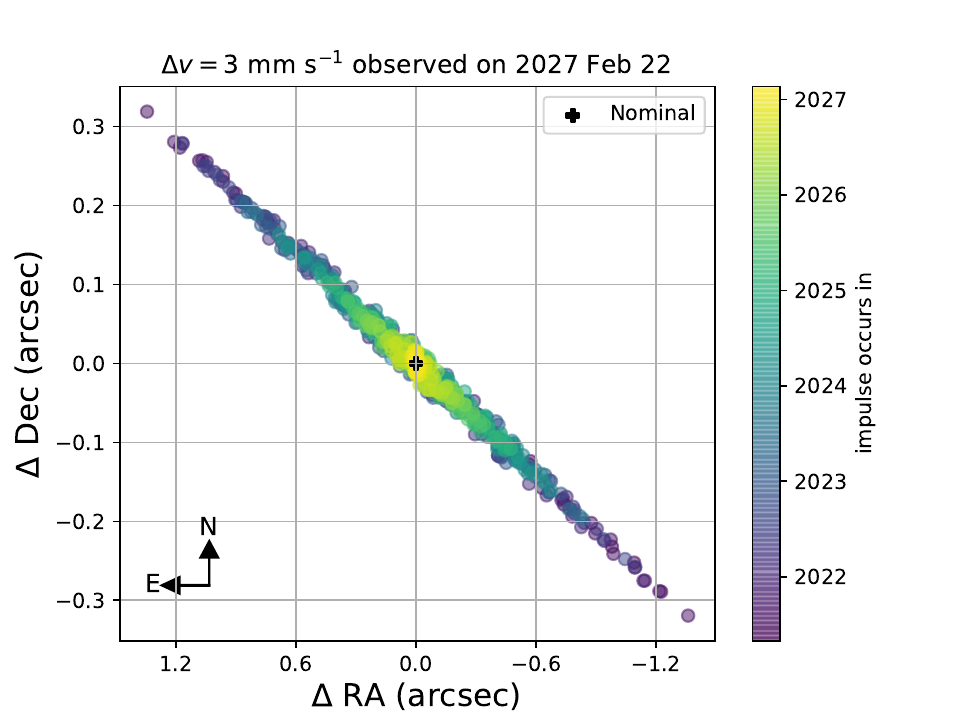}{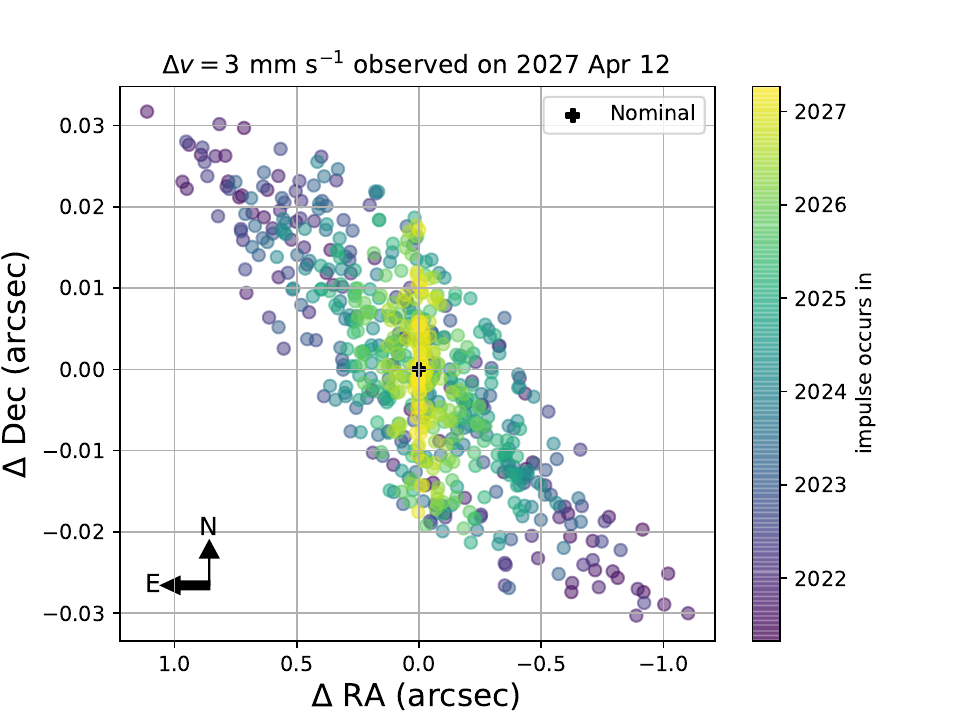}
\plottwo{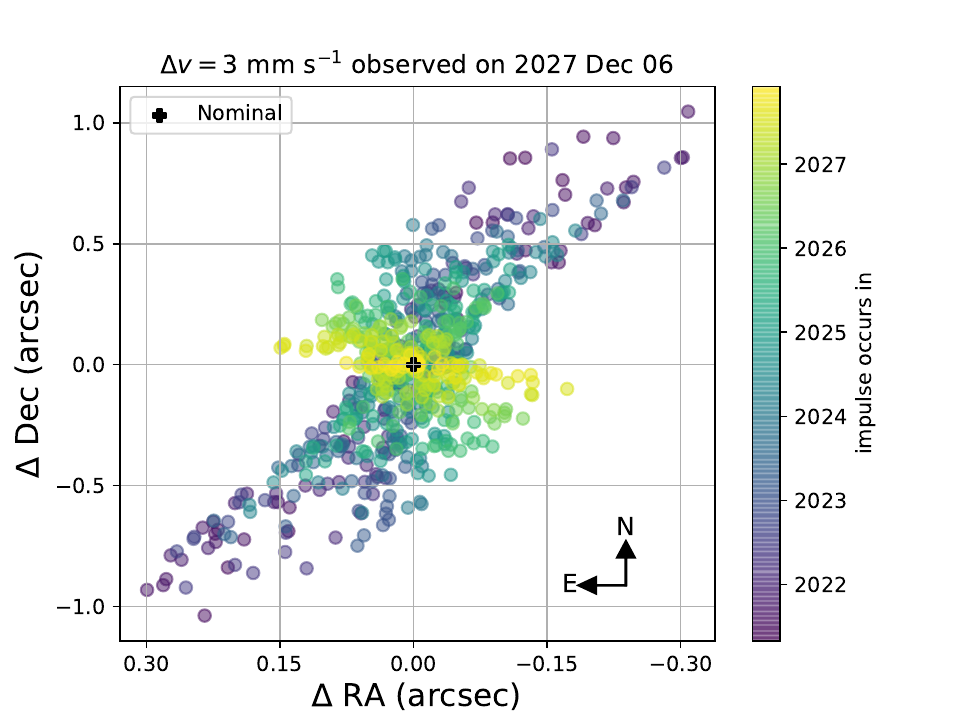}{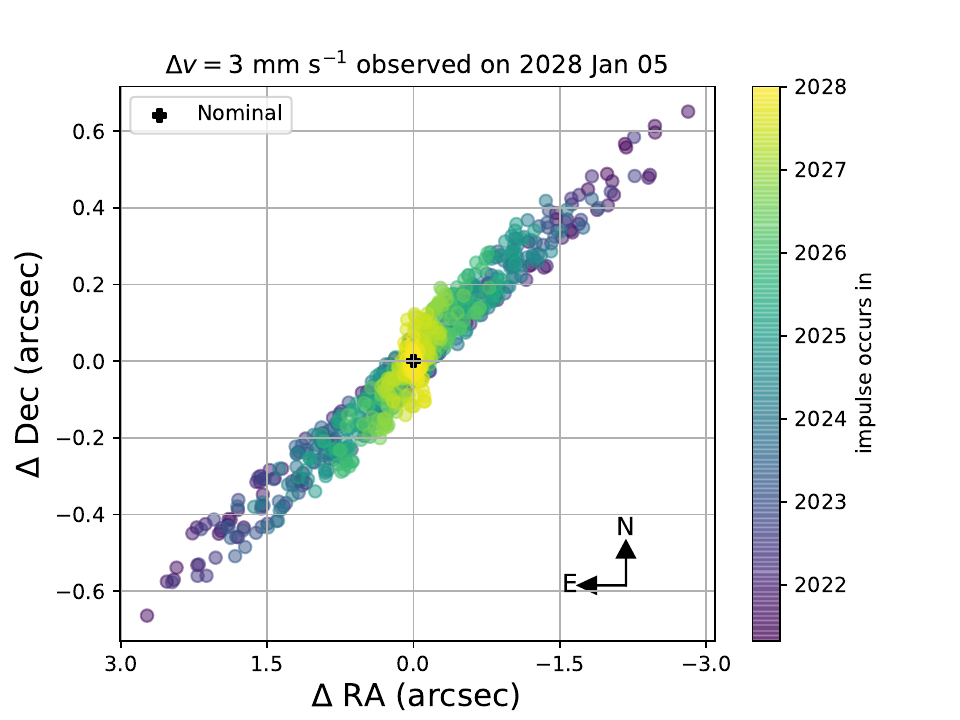}
\caption{The on-sky displacement from the nominal Apophis ephemeris at
  the start (top left) and end (top right) of the early observing
  window, and the start (bottom left) and 30 days into (bottom right)
  of the late observing window, for an impulse $\dv = 3 \times
  10^{-3}$m/s. Displacements are in arcseconds on the sky. The ephemeris uncertainty
  in Apophis' position is less than 0.01 arcseconds in all cases, smaller than the size
  of the plotting symbols.}
\label{fig:onsky-early-keyholes}
\end{figure}

In Section~\ref{sec:sensitivity} we saw that a $\dv$ between $3 \times
10^{-4}$ and $3 \times 10^{-3}$~m/s could move Apophis appreciably
relative to keyholes. The effect of the larger of these $\dv$ on the
on-sky position of Apophis is shown in Figure~\ref{fig:onsky-early-keyholes}.

Figure~\ref{fig:onsky-early-keyholes} illustrates that, if a 1
arcsecond displacement of Apophis is the observational limit, then
many scenarios involving an impulse large enough to move Apophis
appreciably on the target plane might not be detectable from a
single-night observation during either the early or late observing
windows. But displacement $\dtheta$ in on-sky position is not an
unambiguous measure of $\db$. To quantify this more carefully, we ask:
if Apophis is observed to have some displacement $\dtheta$ from its
nominal position, what is the probability that this case is associated
with substantial movement on the target plane?

\begin{figure}[ht!]
\plotone{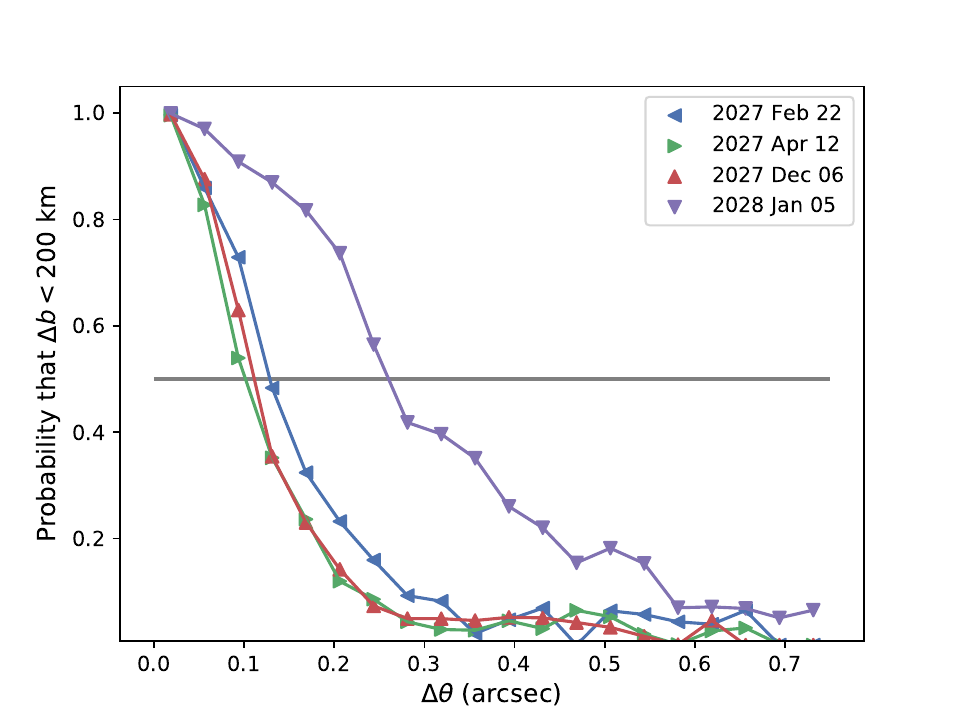}
\caption{The probability that an observed on-sky displacement
  $\dtheta$ from the nominal Apophis ephemeris is associated with a
  target plane displacement $\db < 200$~km, for different dates within
  the early and late observing windows. A high probability indicates
  Apophis is unlikely to have moved substantially with respect to
  important keyholes.  A horizontal line is drawn at a probability of
  0.5.}
\label{fig:dtheta-early-keyholes}
\end{figure}

Figure~\ref{fig:dtheta-early-keyholes} shows the probability of an
observed on-sky offset $\dtheta$ being associated with $\db < 200$~km,
based on all the clones simulated in this work.  Each clone is
weighted by the probability of a small asteroid impact creating its
associated $\dv$, using Eqn~\ref{eq:dv} to determine the asteroid size
$d$ needed, then assuming a small asteroid flux onto Apophis
proportional to that given by \cite{brosparev02} for Earth
(Section~\ref{sec:probability}).  Though this procedure is only
approximate, it provides a useful probability-weighted outlook at
possible outcomes. For example, if a deviation of Apophis from its
nominal ephemeris position of $\dtheta \approx 0.1''$ was measured on
2027 April 12, Figure~\ref{fig:dtheta-early-keyholes} reveals that the
odds are about 50:50 that the case involves substantial motion of
Apophis relative to the keyholes.

From Figure~\ref{fig:dtheta-early-keyholes} we conclude that single
night observations good to 1 arcsecond will be insufficient to
determine if Apophis has moved relative to the important keyholes on
the target plane. However, if Apophis is seen to be displaced by more
than 0.2" from its ephemeris position at this time, then a substantial
displacement of its intersection with the target plane has possibly
occurred. This would not, however, mean that Apophis is on track to
enter a dangerous keyhole; the odds are still in favor of a continued
safe trajectory. But a displacement of Apophis of more than a few
tenths of an arcsecond from nominal in 2027 would be cause for
additional follow-up observations and analysis.

\subsubsection{Earth impact}

The probability of a small asteroid deflecting Apophis onto an
Earth-intersecting orbit in 2029 is exceptionally low, less than 1 in a
billion by the analysis of Section~\ref{sec:probability}. But if this
unlikely event occurs, how will it manifest itself observationally when
Apophis first returns to visibility in 2027?

\begin{figure}[ht!]
\plottwo{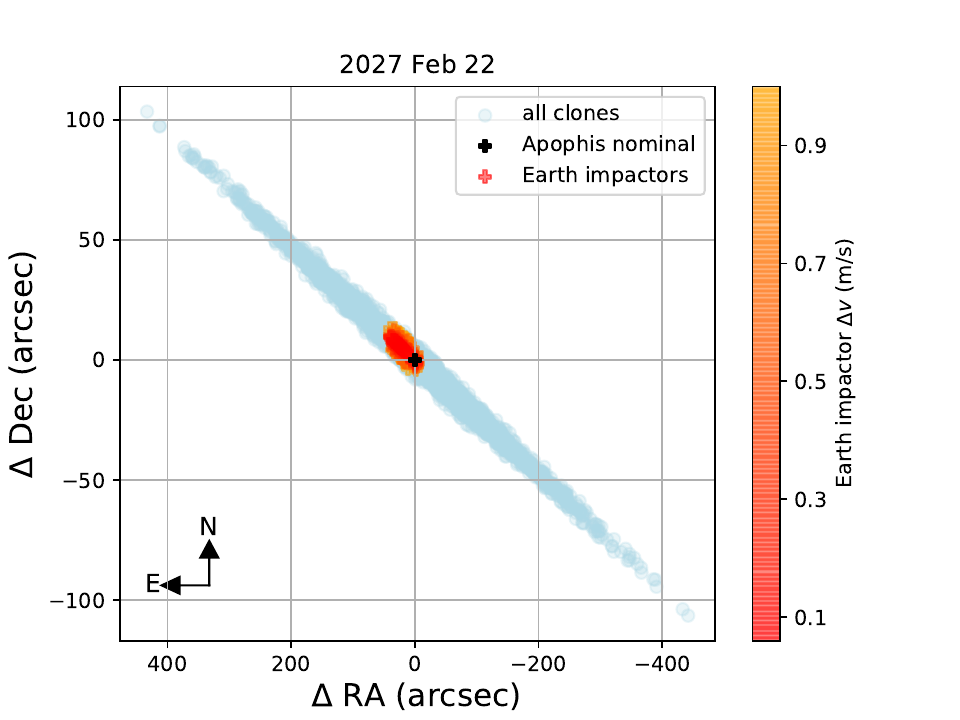}{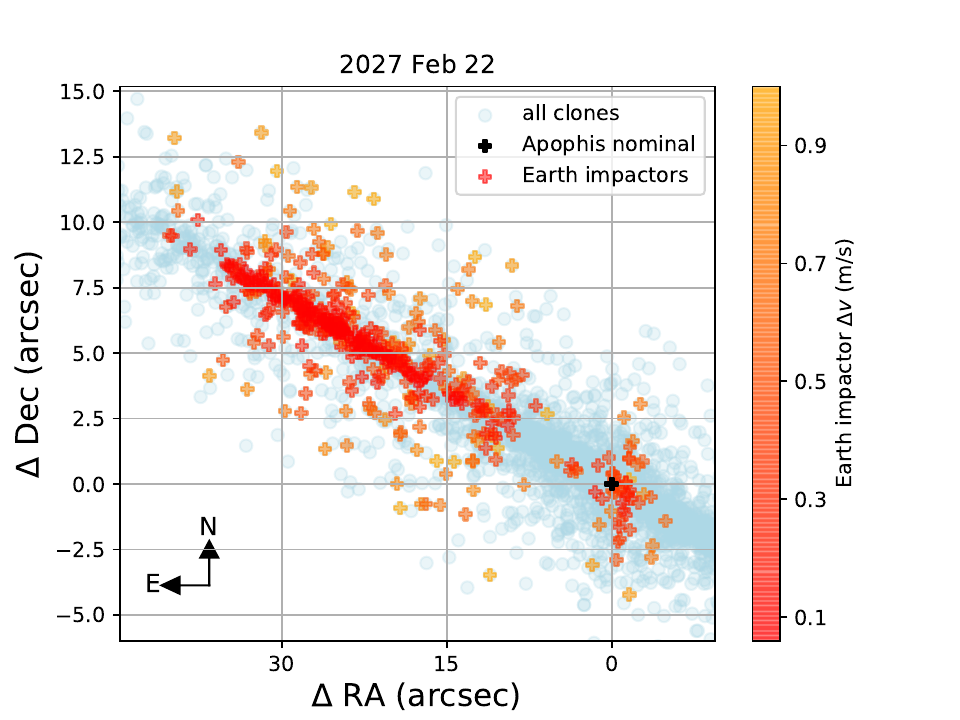}
\plottwo{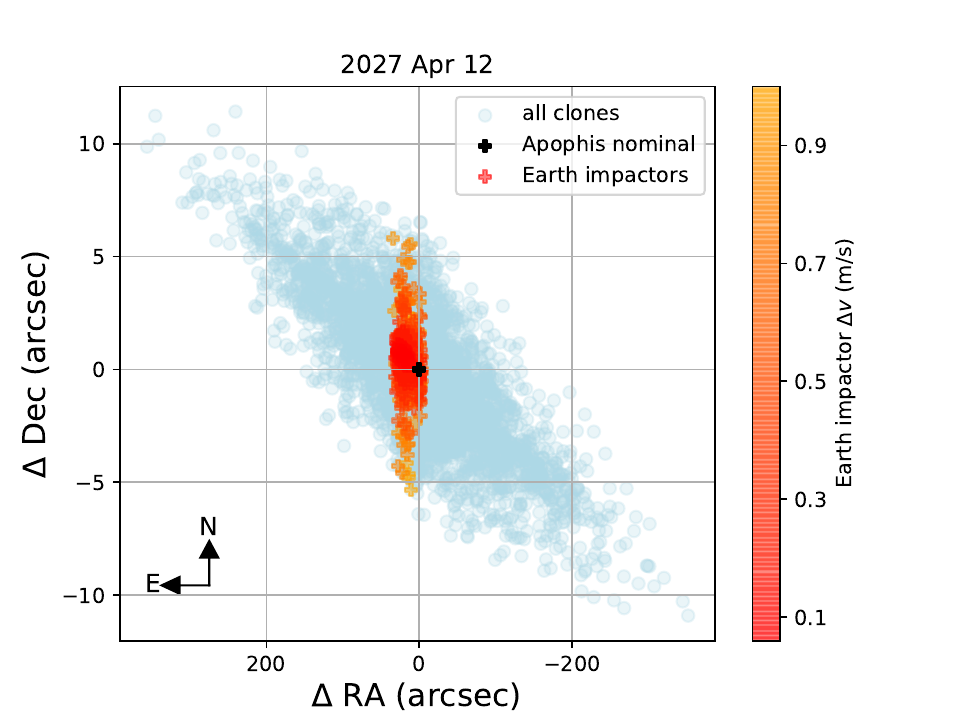}{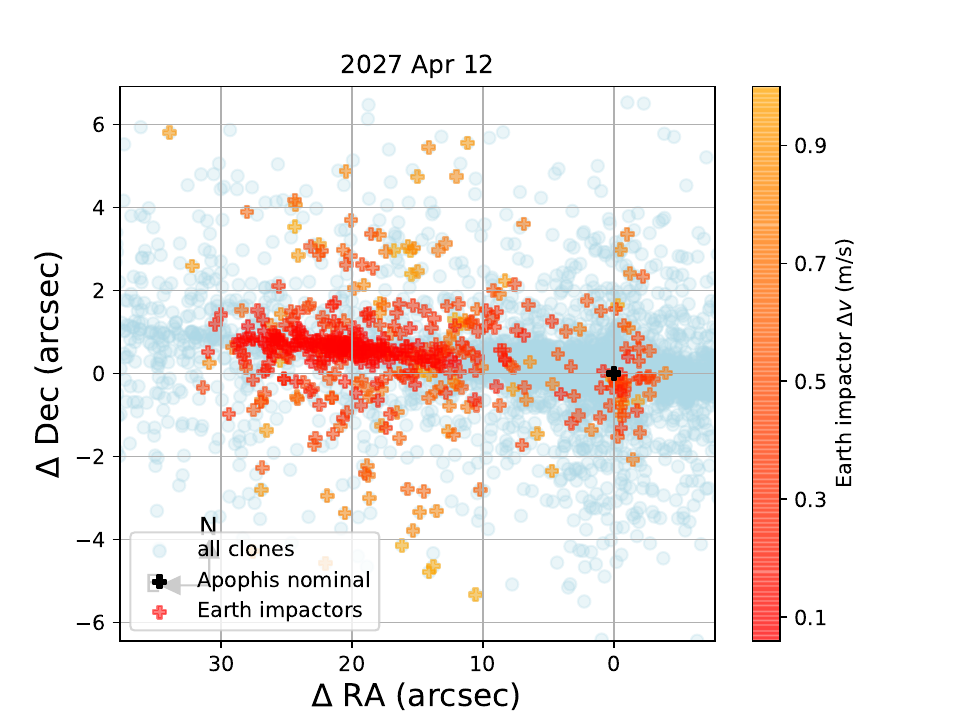}
\caption{The on-sky position of all clones for the beginning (top) and
  end (bottom) of the early observing window, at two different
  magnifications. Clones that would impact Earth are indicated by
  crosses colored by the $\dv$ received.}
\label{fig:onsky-early-impactors}
\end{figure}

Most cases leading to an Earth impact show easily measurable on-sky
position differences from the nominal ephemeris during the early
observing window; these are shown in
Figure~\ref{fig:onsky-early-impactors}. Earth impactors form a
distinct subset; however, positions within the 'red zone' of
Figure~\ref{fig:onsky-early-impactors} are not sufficient to identify
Earth impactors unequivocally, as they overlap with other cases that
have been heavily perturbed, but will not reach our planet.

Though most Earth impactors are significantly displaced from the
nominal on-sky position of Apophis, not all are.  Roughly 1\% of Earth
impactors appear less than 1 arcsecond from the nominal location on
the sky when the early window opens, increasing to 3\% at its
closing. These mostly correspond to cases where the collision that
created the velocity change happened within the 30 days prior to the
date of observation: Apophis has not yet had time to move much from
its nominal trajectory.  Since the dust cloud and/or tail created by
the DART impact was visible for at least 30 days after that event
\citep{karthoyan23, opimursno23}, we expect that the unlikely case of
Apophis being both near its nominal on-sky position and on an
Earth-intercepting trajectory will likely be distinguishable by the
presence of a dust cloud or tail. We do find that a very small fraction of
Earth impactors are located near the nominal on-sky position of Apophis
despite having received an impulse more than 30 days before the
simulated observation, though none more than 180 days.  This suggests
that single-night observations of Apophis are likely to be very
helpful in identifying dangerous scenarios, but may not be sufficient
on their own.

\begin{figure}[ht!]
\plottwo{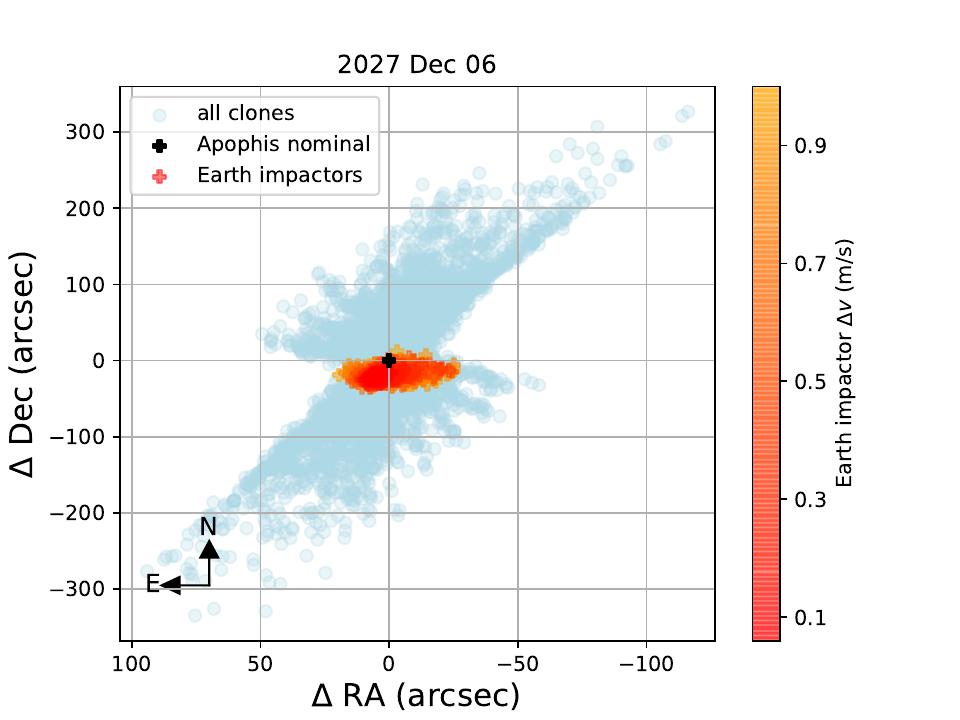}{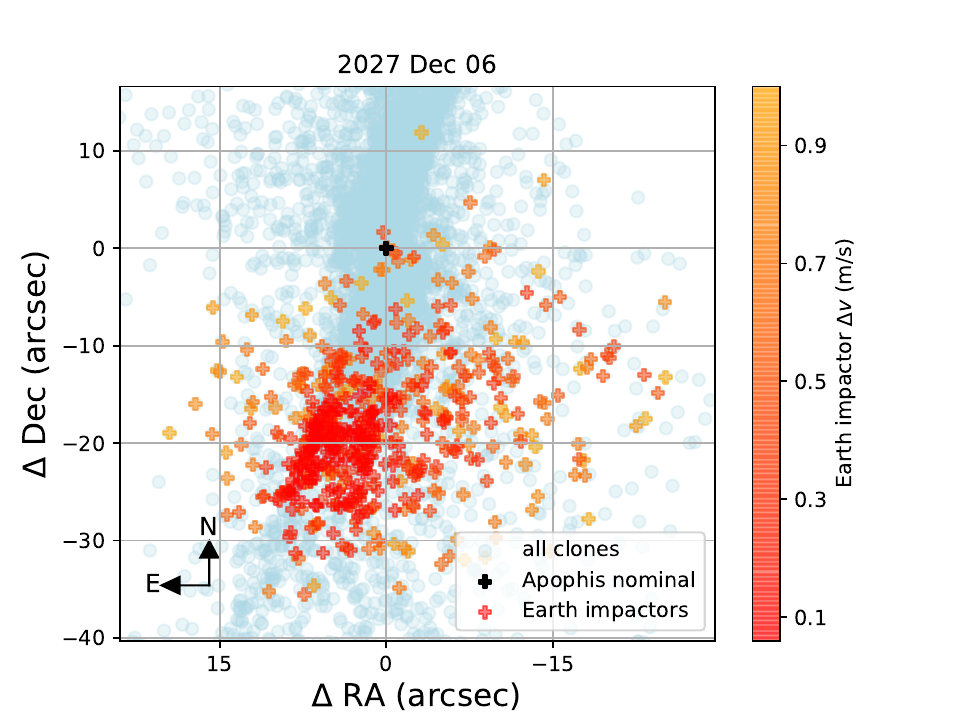}
\plottwo{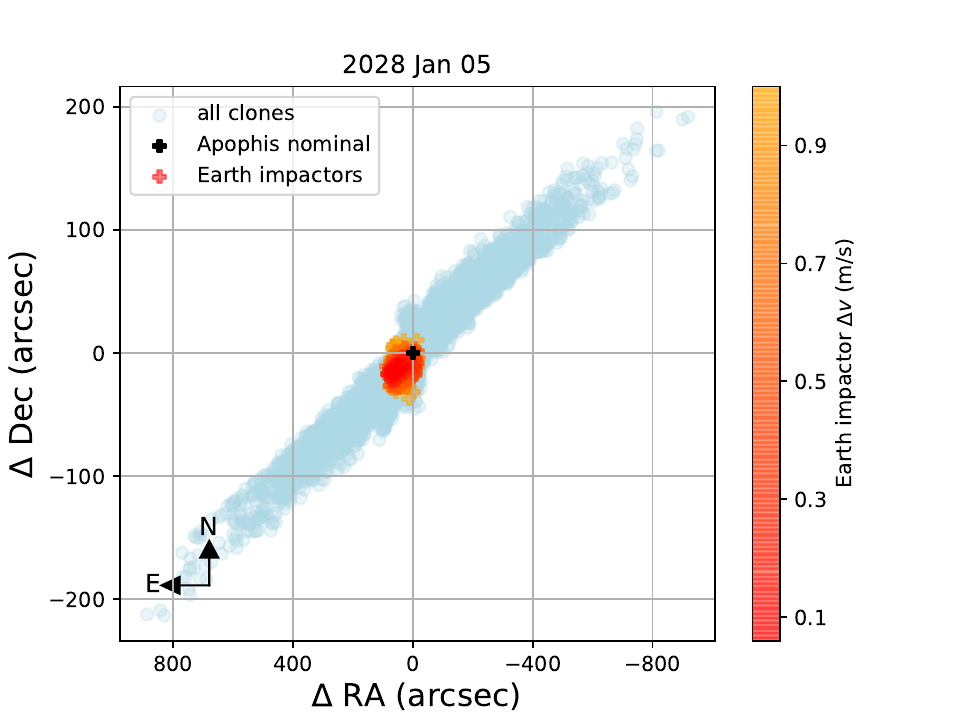}{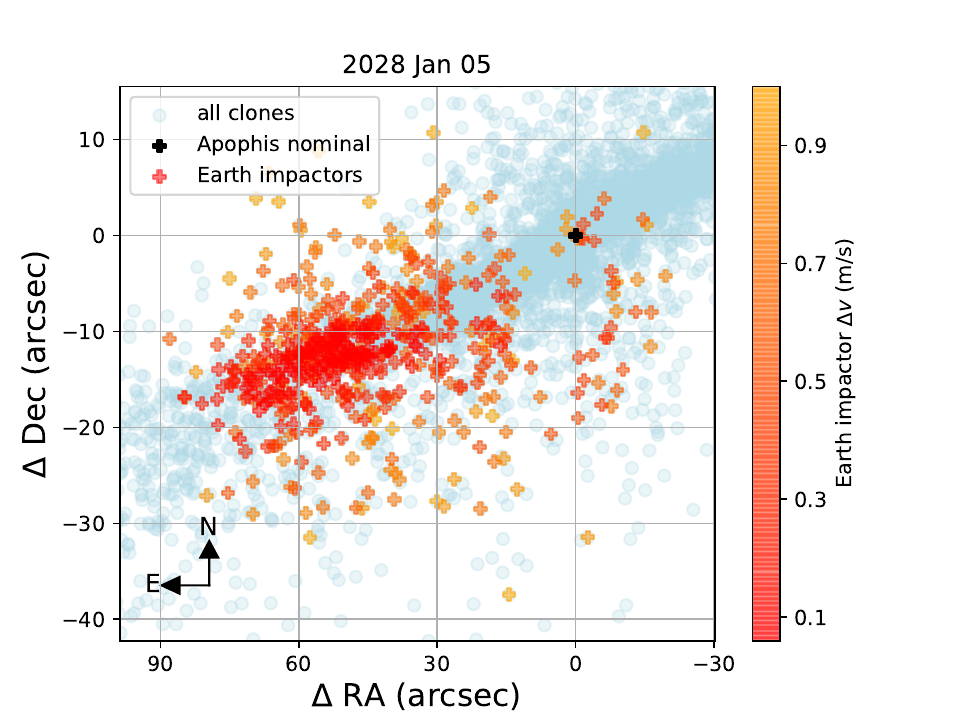}
\caption{The on-sky position of all clones for the beginning of (top) and
  30 days into (bottom) the late observing window, at two different
  magnifications. Clones that would impact Earth are indicated by
  crosses colored by $\dv$.}
\label{fig:onsky-late-impactors}
\end{figure}

During the late observing window, the majority of the Earth-impacting
clones are well away from the nominal ephemeris position. A few of these
cases fall within 1 arcsecond of nominal on the sky
(Figure~\ref{fig:onsky-late-impactors}); however, all such scenarios
examined in this study occured when the impact that deflected Apophis
happened within the 20 days prior to the observation. Given the
persistence of the resulting dust cloud, this case should be
identifiable reliably from telescopic measurements.

\subsubsection{Other observables} \label{sec:magpar}

Since our goal is the rapid assessment of whether or not Apophis might
have undergone a perturbation of concern, we consider briefly whether
other simple observables could be used to assess Apophis' post-2027
orbit.

The apparent magnitude of an asteroid can be used as a proxy for its
distance, if its absolute magnitude is well known. But if a small
asteroid does deflect Apophis, its surface properties and reflectivity
could be changed by the impact. There could also be residual dust in
orbit of Apophis that makes using its apparent magnitude to make
deductions about its trajectory difficult. But a distinct brightening
of Apophis observed in 2027 could be the result of an impact, and
would indicate that further investigation was warranted.

Additional orbital information could be provided by parallax
measurements of distance. Though not traditionally used in orbit
determination, parallax-based techniques are becoming more prevalent
\citep{heimet15,gurottted18,zhashasai22}. Two stations on opposite
sides of the Earth observing Apophis when it emerges from the daytime
sky in 2027 will see it at a distance of approximately 1~au, and so
would measure a parallax $p$
\[
p = \frac{2 R_{E}}{\rm{1~au}} \approx 18~\rm{arcseconds}
\]
where $R_E$ is Earth's radius. The parallax could be used to
determine Apophis' distance if it could be measured precisely enough.
A longer baseline would assist in parallax measurements. The Earth's
orbital motion about the Sun provides roughly 2.5 million km of
baseline per 24 hour period, but parallax measurements would be
complicated by the motion of the asteroid during that period of
time. A space telescope such as the James Webb Space Telescope (JWST)
located at the Earth-Sun L2 point would provide baseline of order 1
million km, much longer than available to Earth observers alone.  JWST
is not designed to point within 85 degrees of the Sun
\citep{nelatcatk04}, but other current or future space telescopes may
be able to contribute to a determination of Apophis' distance via
parallax, in the unlikely event that circumstances warrant it.

\section{Conclusions}

The odds of a small unseen asteroid colliding with Apophis in such a
way as to create a dangerous outcome is exceptionally small. Here we
compute the odds of an impact deflecting it significantly relative to
known keyholes as less than 1 in a million, and of it being deflected
by an amount comparable to its 2029 miss distance as less than 1
in a billion. These probabilities represent only a very low risk, and are below the usual thresholds
considered by asteroid impact warning systems (typically $10^{-6}$,
\cite{milchesan05, roafarche21}).

Apophis will remain largely unobservable until 2027. When it does
return to visibility, a displacement of this asteroid's on-sky
position by more than a few tenths of an arcsecond from nominal during
2027 could indicate that it has been perturbed so as to have 
changed its target $b$-plane intersection substantially relative to
important keyholes. Such a displacement ---if observed--- would not
necessarily indicate that it had been deflected onto a dangerous
trajectory, but would indicate that follow-up and analysis should be
initiated.

For the particularly unlikely case that Apophis gets deflected onto an
Earth-impacting trajectory in 2029, the asteroid's on-sky position in
2027 will differ from the expected ephemeris by tens of arcseconds in
almost all cases. It is however possible that Apophis will appear near
its nominal on-sky position in 2027 despite being on an
Earth-impacting trajectory. These cases correspond to the deflection
occuring just before the observations are taken: Apophis will not yet
have had time to differ much from its nominal trajectory. Such
deflections are found to occur usually within 30 days and no more than
180 days prior for the cases examined here. Such a scenario should be
distinguishable by a residual optical brightening of Apophis caused by
dust generated from the hypervelocity impact that deflected it. We
conclude that the deflection of Apophis by a small asteroid onto a
collision course with Earth in 2029 --- in addition to being
extremely unlikely --- will most likely be quickly eliminated as a
possibility by simple telescopic observations when Apophis returns to
visibility in 2027.

\begin{acknowledgements}
The author thanks A. Christou, P. Brown and two anonymous reviewers
for valuable comments and discussions that much improved this work,
and expresses particular appreciation to observers ---both
professional and amateur--- who continue to contribute to the
worldwide effort of monitoring near-Earth objects.

This work used data from the International Astronomical Union's Minor
Planet Center (MPC). Data from the MPC's database is made freely
available to the public. Funding for this data and the MPC's
operations comes from a NASA PDCO grant (80NSSC22M0024), administered
via a University of Maryland - SAO subaward (106075-Z6415201). The
MPC's computing equipment is funded in part by the above award, and in
part by funding from the Tamkin Foundation.

This work was supported in part by the NASA Meteoroid Environment
Office under Cooperative Agreement No.  80NSSC21M0073, and by the
Natural Sciences and Engineering Research Council of Canada (NSERC)
Discovery Grant program (grant No. RGPIN-2018-05659).
\end{acknowledgements}

\bibliography{Apophis-bib}{}

\bibliographystyle{aasjournal}



\end{document}